\documentclass{pasa}
\usepackage{graphicx}
\usepackage{amsmath}
\usepackage[]{mcode}
\usepackage{lmodern}
\usepackage[authoryear]{natbib}

\title[Algorithm for Gravitational Magnification Maps]
  {A Simple and Practical Algorithm for Accurate Gravitational Magnification Maps}
\author[Walters \& Forbes]
  {S.J. Walters and L.K. Forbes \\
 School of Mathematics and Physics, University of Tasmania, P.O. Box 37, Hobart, 7001, Tasmania, Australia}
\date{Released 2015 Xxxxx XX}

\begin{document}

\begin{abstract}
In this brief communication a new method is outlined for modelling magnification patterns on an observer's plane using a first order approximation to the null geodesic path equations for a point mass lens. For each ray emitted from a source, an explicit calculation is made for the change in position on the observer's plane due to each lens mass. By counting the number of points in each small area of the observer's plane, the magnification at that point can be determined. This allows for a very simple and transparent algorithm. A short \textsc{Matlab} code sample for creating simple magnification maps due to multiple point lenses is included in an appendix.
\end{abstract}


\maketitle

\iftrue
\section{Introduction}

Gravitational lensing is the deflection of light from a distant source by intervening massive objects, resulting in magnification or de-magnification of images. In some regions, areas of high magnification (caustics) are formed. By measuring this magnification, observers passing through these caustics may be able to determine properties of the lens or source. Modelling these phenomena is computationally expensive, particularly for `ray shooting' methods which deflect individual rays, then identify where these rays intersect with the source plane (or the observer's plane), and finally calculate the magnification factor by counting the number of rays which intersect each small area (or pixel) in the plane. Several sophisticated methods have been developed which greatly improve the efficiency of this procedure (for example, see \citet{lew}, \citet{med} and \citet{met}). Also, techniques which make use of parallel processing have been developed to reduce the time required for generating these magnification maps (for example, see  \citet{thom} and \citet{bate}).

In this note, we present an alternative ray shooting method for drawing magnification maps. For each ray emitted by the source, we will derive an expression for its location at the target as the end point of a straight line segment plus a small correction due to each lens in the system. This will be done using a linearized solution to the Schwarzschild path equations, and we will approximate the solution to first order in the small parameter $r_s$, the Schwarzschild radius of the lens. This assumes that $r_s$ is small relative to any other lengths appearing in the equation, principally the distances of closest approach of the ray to the lenses. This linearized solution has been shown to be a good approximation to the full (non-linear) geodesic equations in \citet{wf}.

The present approach does not claim to be faster to run than existing methods. However, it has the advantage that any lens can be easily modelled by simple combination of point masses. The code included in the appendix loops through a list of masses, adding up the perturbation due to each mass. Adding any additional lensing object is merely the inclusion of an item in the list, specifying the object's $x, y$ and $z$ co-ordinates and its mass. The transparency and versatility of this approach will be demonstrated by the modelling of various lens configurations.

\section{Kinematical approach}

In order to apply a fully consistent first order approximation to a ray passing by several lenses, we begin with the first order path equations derived from the acceleration vector for a massless test particle in the gravitational field of a massive lens located at the origin (see \citet{wf2}). These are:

\begin{eqnarray}
\mathbf{\ddot{r}}=\frac{-3 r_s K}{2 r^5}\mathbf{r} \nonumber
\label{schwAcc}
\end{eqnarray}
where $r_s$ is the Schwarzschild radius of the lens, $K$ is the square of the impact parameter, and $\mathbf{r}$ is the normal position vector, and $r=\vert\mathbf{r}\vert$ is the distance from the origin. The dot-notation refers to differentiation by the proper time parameter, $\tau$. Using a first order expansion in $r_s$, we approximate the light path as

\begin{eqnarray}
x&=&X_0+r_s X_1+O(r_s^2) \nonumber \\
y&=&Y_0+r_s Y_1+O(r_s^2) \nonumber \\
z&=&Z_0+r_s Z_1+O(r_s^2),
\label{xyz1}
\end{eqnarray}
where the zeroth order components (straight lines) are given by
\begin{eqnarray}
X_0&=&C_1 \tau+C_2 \nonumber \\
Y_0&=&C_3 \tau+C_4 \nonumber \\
Z_0&=&C_5 \tau+C_6, \nonumber
\label{xyz0}
\end{eqnarray}
and the first order corrections are
\begin{eqnarray}
X_1&=&\frac{X_0}{2 R_0}-\frac{R_0}{K_0}(C_2-B C_1)+C_{11} \tau + C_{21} \nonumber \\
Y_1&=&\frac{Y_0}{2 R_0}-\frac{R_0}{K_0}(C_4-B C_3)+C_{31} \tau + C_{41} \nonumber \\
Z_1&=&\frac{Z_0}{2 R_0}-\frac{R_0}{K_0}(C_6-B C_5)+C_{51} \tau + C_{61}.
\label{XYZ1}
\end{eqnarray}

In these equations, $R_0(\tau)$ is the distance at `time' $\tau$ from the lens mass to the point where the test particle would be if it were not deflected (that is, in a straight line path from the source). The constants $C_1$ to $C_6$ are determined by initial conditions, and the constant $B=C_1 C_2 +C_3 C_4 + C_5 C_6$. These equations are for a lens mass located at the origin. In the current study, it is convenient to place the source, rather than the lens, at the origin, as this will allow us easily to include more than one lens mass. We will specify that the ray leaves the origin at $\tau=0$. In this case, $C_2=C_4=C_6=0$, and the zeroth order terms simplify to

\begin{eqnarray}
X_0&=&C_1 \tau \nonumber \\
Y_0&=&C_3 \tau \nonumber \\
Z_0&=&C_5 \tau, \nonumber
\label{xyz0b}
\end{eqnarray}
and the first order corrections are now
\begin{eqnarray}
X_1&=&\frac{X_0-x_m}{2 R_0}+\frac{R_0}{K_m}(x_m+B_m C_1)+C_{11} \tau + C_{21} \nonumber \\
Y_1&=&\frac{Y_0-y_m}{2 R_0}+\frac{R_0}{K_m}(y_m+B_m C_3)+C_{31} \tau + C_{41} \nonumber \\
Z_1&=&\frac{Z_0-z_m}{2 R_0}+\frac{R_0}{K_m}(z_m+B_m C_5)+C_{51} \tau + C_{61}.
\label{XYZ1b}
\end{eqnarray}
The new constants $x_m, y_m$ and $z_m$ are the co-ordinates of the massive lens (with subscript $m$ for mass), and the other terms relating to this mass are given by
\begin{eqnarray}
R_0&=&\sqrt{\tau^2 + R_m^2 + 2 B_m \tau} \nonumber \\
R_m&=&\sqrt{x_m^2+y_m^2+z_m^2} \nonumber \\
B_m&=&-(C_1 x_m+C_3 y_m+C_5 z_m) \nonumber \\
K_m&=&R_m^2-B_m^2. \nonumber
\label{ra}
\end{eqnarray}
As we have specified that the light ray leaves the origin at $\tau=0$, Equations (\ref{XYZ1b}) are zero at that time. Solving defines three of the constants as:
\begin{eqnarray}
C_{21}&=&\frac{x_m}{2 R_m}-\frac{R_m(x_m + C_1 B_m)}{K_m} \nonumber \\
C_{41}&=&\frac{y_m}{2 R_m}-\frac{R_m(y_m + C_3 B_m)}{K_m} \nonumber \\
C_{61}&=&\frac{z_m}{2 R_m}-\frac{R_m(z_m + C_5 B_m)}{K_m}. \nonumber
\label{c214161}
\end{eqnarray}

By also specifying that the ray leaves the origin at some trajectory $(\phi,\theta)$, where $\phi$ is the azimuthal angle and $\theta$ is the inclination angle above the $x-y$ plane, along with the speed constraint equation, $C_1 C_{11}+C_3 C_{31}+C_5 C_{51}=0$ (see \citet{wf2}), we can solve for the other constants as follows:
\begin{eqnarray}
C_{1}&=&\cos \phi \cos \theta \nonumber \\
C_{3}&=&\sin \phi \cos \theta \nonumber \\
C_{5}&=&\sin \theta. \nonumber \\
C_{11}&=&-(C_1 B_m + x_m)\frac{B_m}{R_m} \big(\frac{1}{K_m}+\frac{1}{2 R_m^2}\big) \nonumber \\
C_{31}&=&-(C_3 B_m + y_m)\frac{B_m}{R_m} \big(\frac{1}{K_m}+\frac{1}{2 R_m^2}\big) \nonumber \\
C_{51}&=&-(C_5 B_m + z_m)\frac{B_m}{R_m} \big(\frac{1}{K_m}+\frac{1}{2 R_m^2}\big) \nonumber
\label{c123}
\end{eqnarray}

Substituting these constants into the first order corrections (\ref{XYZ1}) and then into the path equations (\ref{xyz1}) we have the general path equations for rays leaving the origin at $\tau=0$ in a system containing a single lensing object:

\begin{eqnarray}
x&=&C_1\tau+\frac{r_s}{2}\bigg[C_1\bigg(\frac{\tau}{R_0}-\frac{B_m}{K_m}P\bigg)+x_m Q\bigg]+O(r_s^2) \nonumber \\
y&=&C_3\tau+\frac{r_s}{2}\bigg[C_3\bigg(\frac{\tau}{R_0}-\frac{B_m}{K_m}P\bigg)+y_m Q\bigg]+O(r_s^2) \nonumber \\
z&=&C_5\tau+\frac{r_s}{2}\bigg[C_5\bigg(\frac{\tau}{R_0}-\frac{B_m}{K_m}P\bigg)+z_m Q\bigg]+O(r_s^2) \nonumber
\end{eqnarray}

where $P=\frac{B_m \tau}{R_m}(3-\frac{B_m^2}{R_m^2})+2R_m-2R_0$ and $Q=\frac{1}{R_m}-\frac{1}{R_0}-\frac{P}{K_m}$ have been introduced for readability.

The correction to the straight line path is given above, accurate to first order in the small variable $r_s$, which serves as a surrogate for the mass of the lensing object, since $r_s=2 G M/c^2$ where $M$ is the mass of the lens, $G$ is Newton's constant and $c$ is the speed of light. In this paper we are using geometrized units, that is $G=c=1$, unless noted otherwise. The correction due to additional lensing masses may now be easily included. As the approach undertaken here has been one of linearisation of the path equations, the superposition principle holds, and the change to the path is the sum of corrections for each mass. To clarify, this comes about by making the assumption that summation of acceleration components will be accurate at least to first order in $r_s$. Explicitly, we are saying that the acceleration due to $n$ massive objects is:
\begin{eqnarray}
\mathbf{\ddot{r}}=\sum\limits_{i=1}^n \frac{-3 (r_s)_i K_i}{2 r_i^5}\mathbf{r_i}
\label{schw1}
\end{eqnarray}
where the sum is over all the massive bodies in the system, each with its own value for mass ($r_s$), square of the angular momentum of the light ray about the mass ($K$) and position of the particle relative to the mass ($\mathbf{r}$).

We now have the first order path equations for $x, y$ and $z$. This enables us to solve for $\tau$ to first order in $r_s$, for various observer locations. We can then solve the equations for that value of $\tau$ to determine the $x, y$ and $z$ values at the observer's location. This procedure will be illustrated in this paper by solving for an observer sphere and for an observer plane.

\subsection{Ray intersections with an observer's sphere}\label{sectionsphere}
In order to evaluate the intersection of light rays with some outer sphere, centred on the source, we specify that the ray meets a sphere of radius $R$ at some `time' $\tau$.  That is, let $\tau=T_0+r_s T_1+O(r_s^2)$, and solve the following equation for $T_0$ and $T_1$:
\begin{eqnarray}
R^2&=&x^2+y^2+z^2+O(r_s^2). \nonumber
\label{rsqr}
\end{eqnarray}
The solution of this equation leads to the following zeroth-order and first-order components of $\tau$:
\begin{eqnarray}
T_0&=&R \nonumber \\
T_1&=&\frac{B_m}{R_m}-\frac{R + B_m}{R_f}, \nonumber
\label{t1}
\end{eqnarray}
where $R_f=\sqrt{R^2+R_m^2+2 B_m R}$ is the distance from the lens to the observer. Substituting in this value of $\tau$, the values of $x, y$ and $z$ at the sphere can be calculated, given an initial trajectory ($\phi, \theta$). After some simplification, these final values may be written:
\begin{eqnarray}
x_f&=&C_1 R+\frac{r_s}{2}(x_m + C_1 B_m) F+O(r_s^2) \nonumber \\
y_f&=&C_3 R+\frac{r_s}{2}(y_m + C_3 B_m) F+O(r_s^2) \nonumber \\
z_f&=&C_5 R+\frac{r_s}{2}(z_m + C_5 B_m) F+O(r_s^2) \nonumber
\label{xyzf}
\end{eqnarray}
in which
\begin{eqnarray}
F = \frac{1}{R_m}-\frac{1}{R_f} -2\frac{R_m-R_f}{K_m}-\frac{R B_m}{K_m R_m}(3 - \frac{B_m^2}{R_m^2}) \nonumber.
\label{f}
\end{eqnarray}

We are now in a position to plot the points where rays from a star at the centre of an observer's sphere intersect with that sphere. Fig. \ref{fig1} shows the points due to a central source, with a massive lens ($r_s=1$) located at (20,0,0), and a secondary lens ($r_s=0.1$) located at (20,11.8,0). For clarity, the rear part of the sphere has not been plotted.

\begin{figure}
\vspace{1cm}
\caption{Caustic pattern on the surface of a sphere due to a binary lens. The secondary object has one tenth the mass of the primary.}
\includegraphics[width=\columnwidth,height=6cm]{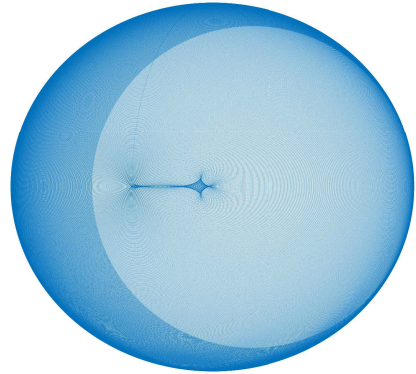} \\
\label{fig1}
\end{figure}

\subsection{Magnification on an observer's plane}\label{sectionplane}
Solving for the magnification on a plane surface follows a similar procedure. In this case, we will solve for a plane which is perpendicular to the $x$-axis. This does not involve any loss of generality, as the axes can be arbitrarily positioned, each lensing object having its own $x, y$ and $z$ co-ordinates. Setting the final $x$-position of the ray to a constant, $X_f$, we solve for $\tau$ and then $y$ and $z$ in a similar way to that in section \ref{sectionsphere} above to obtain:

\begin{eqnarray}
	Y_{f} = \frac{C_3 X_f}{C_1}+\frac{r_s F}{2}(y_m-\frac{C_3 x_m}{C_1}) \nonumber \\
	Z_{f} = \frac{C_5 X_f}{C_1}+\frac{r_s F}{2}(z_m-\frac{C_5 x_m}{C_1})
\label{yz2}
\end{eqnarray}
where 
\begin{eqnarray}
F&=&\frac{1}{R_m}-\frac{1}{R_c}+2\frac{R_c-R_m}{K_m}-2\frac{T_0 B_m}{K_m R_m}-\frac{T_0 B_m}{R_m^3} \nonumber \\
T_0&=&\frac{X_f}{C_1} \nonumber \\
R_c&=&\sqrt{T_0^2 + R_m^2 + 2 B_m T_0}
\label{f2}
\end{eqnarray}

Using these formulae, points can be plotted on the observer's plane. Beginning with an array of regularly spaced rays leaving the source, the final position is calculated. A magnification map can be produced by counting the number of points within each small area of the observer's plane. In Fig. \ref{fig2}, a source is placed at the origin, a primary mass with $r_s=0.01$ is at (20,0,0) and a secondary mass is at (20,0.9,0) with $r_s=10^{-6}$. The intersection of each ray with a plane at $x=2000$ has been calculated and a smooth binning routine \citep{per} has been used to assign a colour based on number of points in each bin.

To simulate a many body system, 16 masses have been placed near $(20,4,0)$, using a random number generator to ensure their $y$ co-ordinates lay in the interval $4<y<6$, and their $z$ co-ordinates lay in $-0.75<z<0.75$, as may be seen from the \textsc{Matlab} code in the appendix. The magnification map resulting from the code provided in the appendix is shown in Fig. \ref{fig3}. It clearly consists of many caustic patterns, due to the 16 masses.

\begin{figure}
\vspace{1cm}
\caption{Caustic pattern on a plane due to the lensing action of a planetary system. The star's mass is 10000 times that of the planet.}
\includegraphics[width=\columnwidth]{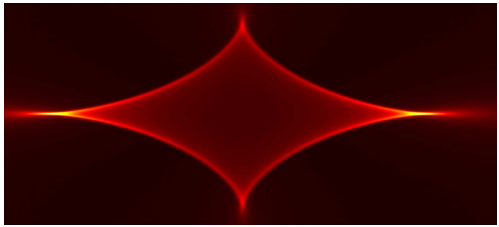} \\
\label{fig2}
\end{figure}

\begin{figure}
\vspace{1cm}
\caption{Caustic pattern on a plane due to the lensing action of 16 masses. The code for this plot is included in the appendix.}
\includegraphics[width=\columnwidth]{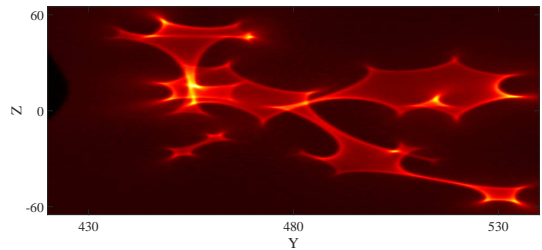} \\
\label{fig3}
\end{figure}

A simulation of an elliptical galaxy has been developed to produce the caustic plot in Fig. \ref{fig4}. Two hundred thousand masses have been placed randomly in an elliptical structure with an inverse squared density. The elliptical shape breaks the spherical symmetry and produces the diamond-shaped caustic pattern. For this figure the lens approximates the central bar of the lensing galaxy in the Einstein Cross (\citet{huch}, \citet{wam}). The lens-source and lens-observer distances have also been set to correspond to the Einstein Cross. With the point source at the origin, the lensing galaxy is placed at 7600, and the observer's plane is at 8000. All distances are in millions of light years. The lensing galaxy has a mass of $1.5 \times 10^{10}$ solar masses, and a height of $0.34$ arcminutes as seen from the observer's plane by an observer on the optical axis. A path through the caustic has been chosen, and the corresponding light curve is shown below the magnification map. In the lower image, a structure of $10^4$ solar masses has been added. The small deviation can be seen in the magnification map and in the corresponding light curve below it.

Computational time for this procedure is proportional to the number of rays multiplied by the number of lensing masses. For the elliptical galaxy with $200 000$ stars, an array of $4 \times 10^6$ light rays was used. The running time for each of these maps on a desktop pc (Core I7) with four cores was $6$ hours.

\begin{figure}
\vspace{1cm}
\caption{Caustic pattern on a plane due to the lensing action of 200 000 masses placed randomly in an elliptical structure with an inverse squared density, designed to roughly approximate the central bar of the lensing galaxy in the Einstein Cross. The total mass of this galaxy is $1.5 \times 10^{10}$ solar masses. Distances are in millions of light years. In the lower half of the figure, an additional substructure of $10^4$ solar masses has been added. The magnification map and corresponding light curve show a small deviation due to this substructure.}
\includegraphics[width=\columnwidth]{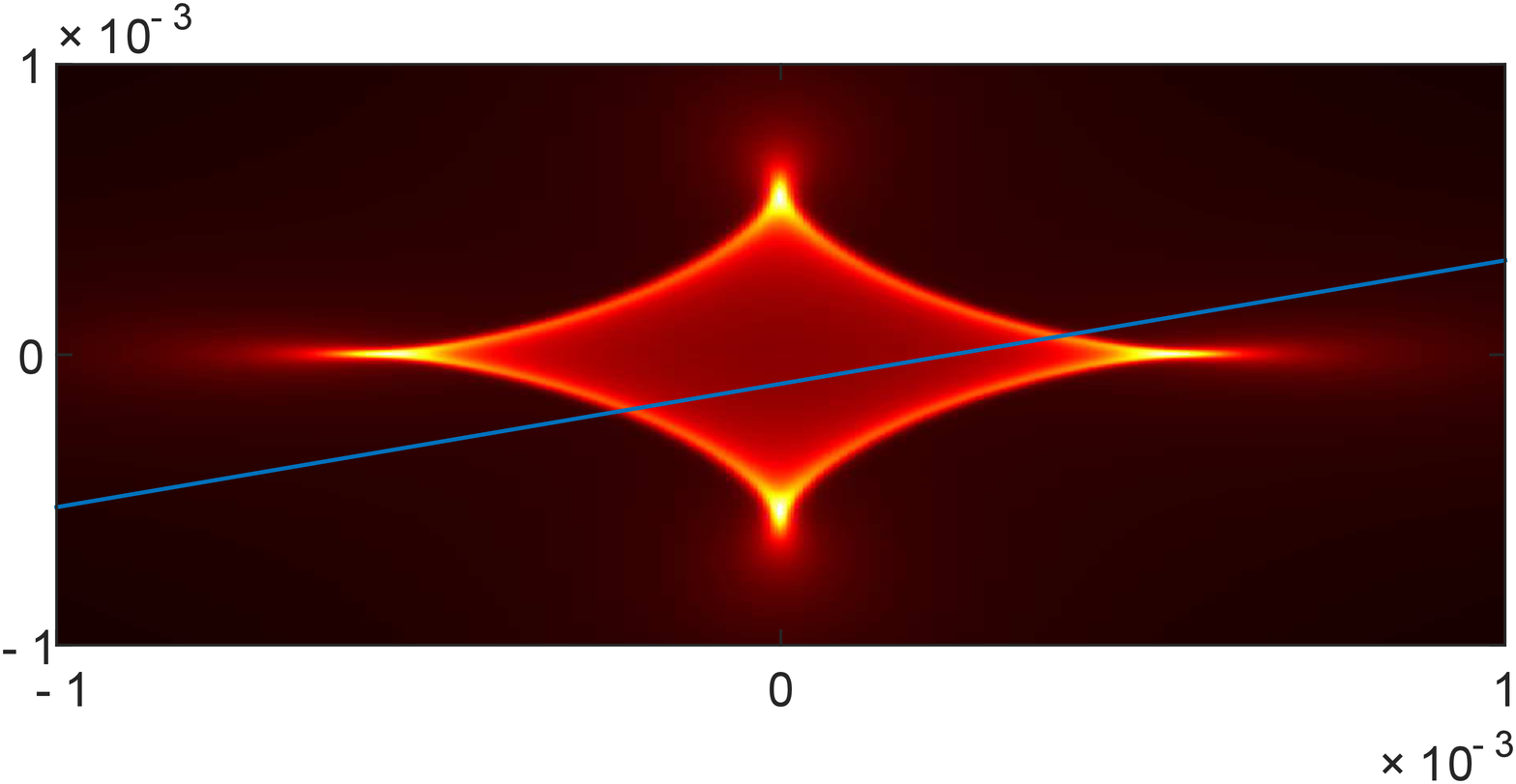} \\
\includegraphics[width=\columnwidth]{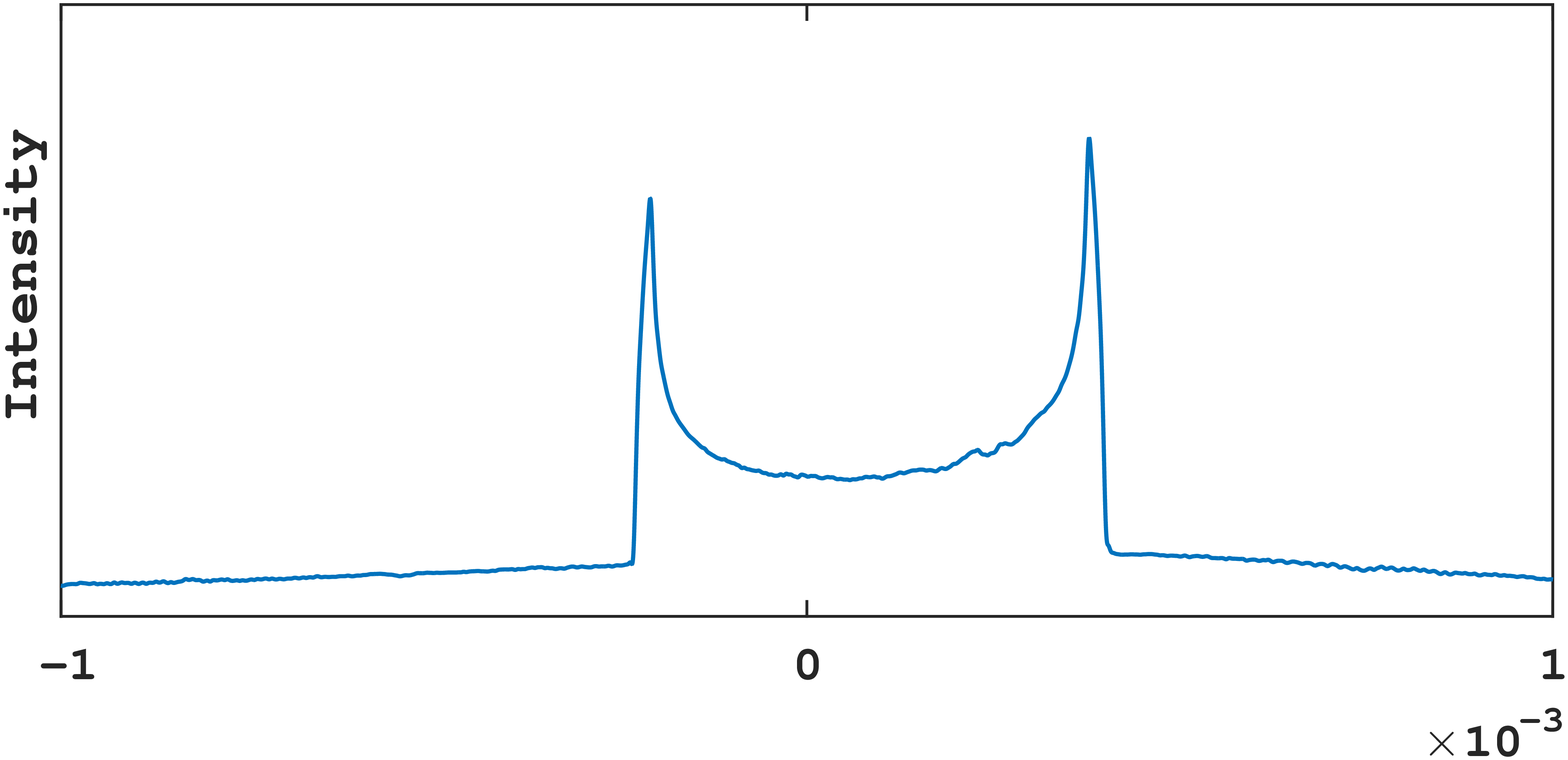} \\
\includegraphics[width=\columnwidth]{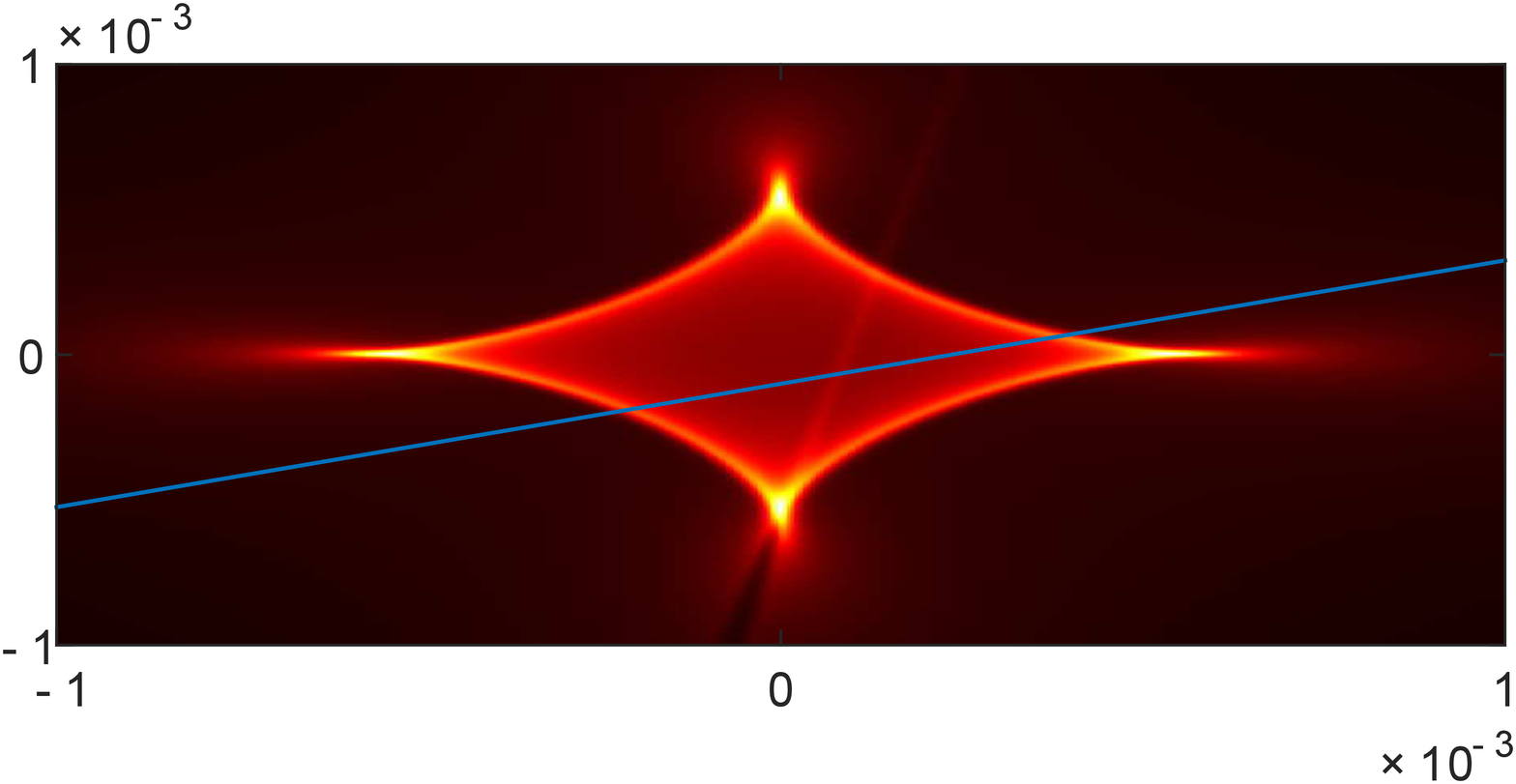} \\
\includegraphics[width=\columnwidth]{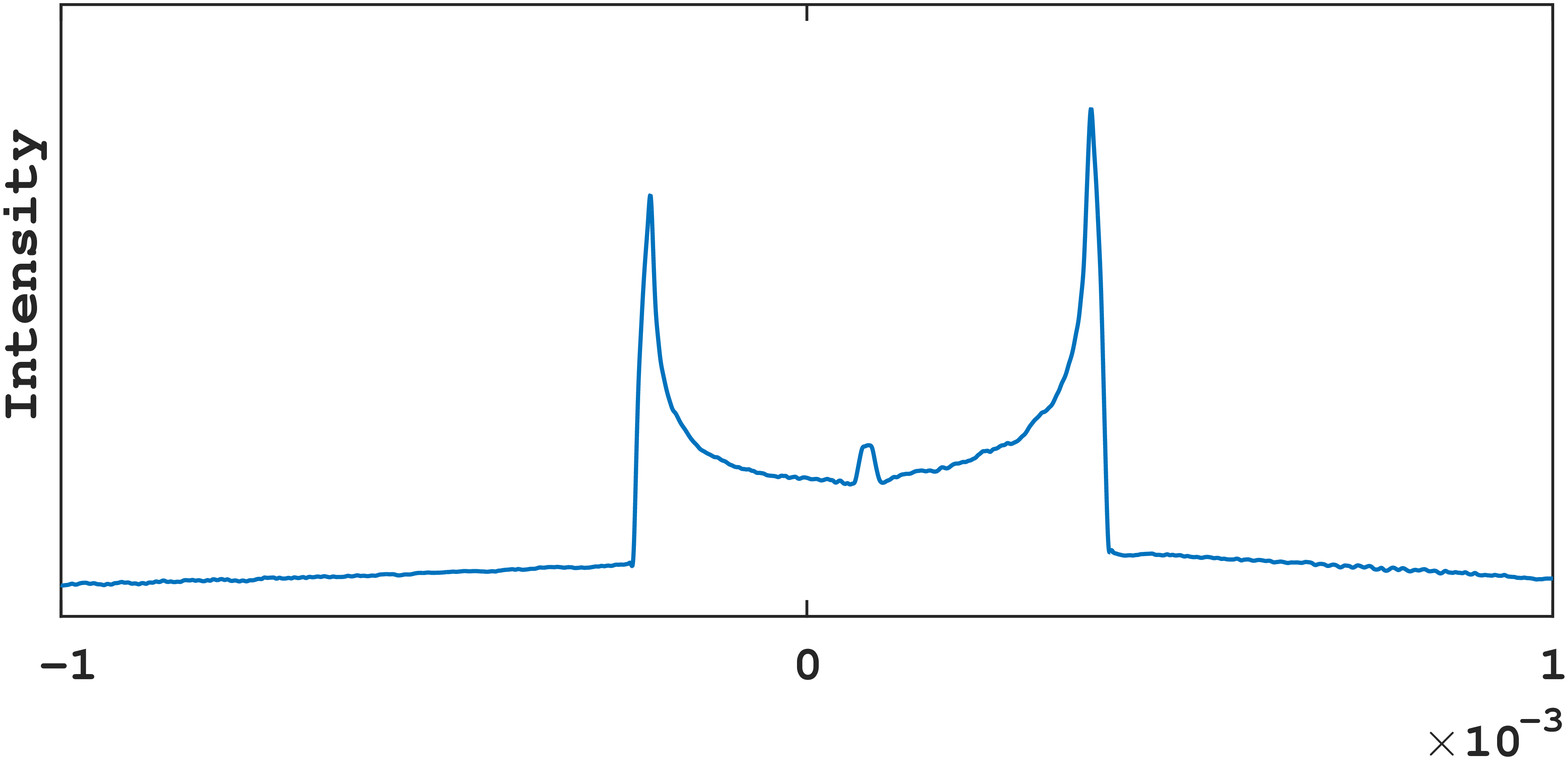} \\
\label{fig4}
\end{figure}

\section{Conclusion}

This short paper has presented a new method for generating magnification patterns due to the gravitational effects of massive objects on light rays from a source. An algorithm has been described for solving the linearised first order path equations for rays from the origin passing through a lens and intersecting with a viewing surface. This general algorithm has been illustrated by solutions for two simple viewing surfaces, a sphere surrounding the source, and a plane.

Once the solution has been derived for the linearised path equations, the final locations for an initial array of light rays can be found by first calculating the value of the time parameter at the observer's location, and from that, the small deviations due to each mass in the system can be obtained. As the path equations have been linearised, the superposition principle holds and the total deflection can be found as the sum of the deflections due to each of the massive objects. This makes it simple to add many lensing objects to the system. It is hoped that the method and code presented here may be helpful for quickly identifying the magnification patterns for any desired lens configuration.

An interesting area of future research would be to solve the equations for incoming angle at the observer's position rather than for position over the observer's plane. This would allow a modelling of macro-lensing effects, and reproduction of the observer's view of the Einstein Cross at a single point, rather than of the spatially extended magnification map.

A further modification would be to generate light curves which reflect the motion of the lensing objects over the time period, in the case where there is significant change in configuration of the lens during the traversal of the magnification map. In the approach covered in this note, this presents a significant challenge, as it involves continually creating new sections of the magnification map as the lens configuration changes.

\section*{Acknowledgements}
The authors gratefully acknowledge the financial support for this work provided by Australian Research Council Discovery Grant DP140100094.

The authors are also indebted to Dr L. Wyrzykowski for helpful comments on an earlier draft of this paper.

%\section*{Bibliography}

\section{Appendix}
The code presented here is a very basic implementation of the planar solution described in the text.
\begin{lstlisting}
%set up an array of rays large enough to cover the
%target area. This rectangular array is a suitable
%approximation for an isotropically emitting
%source near the optical axis, but more care is
%required if plotting further out.
xf=2000;  %observer plane at x=2000
yf0=390;
yf2=590;
zf0=-100;
zf2=105;
gy=500;  %number of points in y-direction
gz=500;  %number of points in z-direction
yf=linspace(yf0,yf2,gy);
zf=linspace(zf0,zf2,gz);

%determine target location for magnification map
yt0=420;
yt2=540;
zt0=-65;
zt2=65;

%set up meshgrid of rays and their initial
%velocity components
y=ndgrid(yf,zf)';
z=ndgrid(zf,yf);
R=sqrt((xf^2+y.^2+z.^2));
c3=y./R;
c5=z./R;
c1=sqrt(1-c3.^2-c5.^2);
%set up array of lensing object locations and
%masses (x,y,z,mass)
masses=[
   20.0000    4.3334   -0.5146    0.0001
   20.0000    4.2860    0.2038    0.0007
   20.0000    4.2442   -0.0532    0.0005
   20.0000    4.7654    0.7264    0.0002
   20.0000    4.7766   -0.0858    0.0008
   20.0000    4.3825    0.7482    0.0008
   20.0000    5.0801   -0.6292    0.0002
   20.0000    4.5315    0.1704    0.0008
   20.0000    5.0130   -0.4282    0.0003
   20.0000    5.0609    0.2037    0.0004
   20.0000    4.9703   -0.5343    0.0003
   20.0000    5.5696    0.0963    0.0009
   20.0000    5.4713   -0.7395    0.0007
   20.0000    4.3961    0.5765    0.0002
   20.0000    4.4599   -0.4150    0.0001
   20.0000    5.3170    0.2771    0.0009];
[n,~]=size(masses); %number of lensing masses
%pre-allocate space for storing first order
%corrections
yrs=zeros(size(y));
zrs=yrs;
%calculate first order deviations due to each mass
for i=1:n;
    xn=masses(i,1);
    yn=masses(i,2);
    zn=masses(i,3);
    mn=masses(i,4);
    
    Rn =sqrt(xn^2 + yn^2 + zn^2);
    Bn = -(c1*xn+c3*yn+c5*zn);
    Kn = Rn^2 - Bn.^2;
    Rf = sqrt(xf^2+c1.^2*Rn^2+2*xf*c1.*Bn);
    F = 1/Rn-c1./Rf-(2*Rn)./Kn+(2*Rf)./(c1.*Kn)
      -(2*xf*Bn)./(c1.*Kn*Rn)-(xf*Bn)./(c1*Rn^3);
    
    yrs=yrs+mn/2*(yn-c3*xn./c1).*F;
    zrs=zrs+mn/2*(zn-c5*xn./c1).*F;
end
%final locations equal straight line plus sum of
%corrections
yy=c3*xf./c1+yrs; 
zz=c5*xf./c1+zrs;
%sort the data points into two columns for y and z
%for plotting
data=reshape(cat(3,yy,zz),[gy*gz,2,1]);
%for display, only keep the points falling within
%the desired target area
datanew=[];
for k=1:gy*gz
    if data(k,1)>yt0
        if data(k,1)<yt2
            if data(k,2)>zt0
                if data(k,2)<zt2
                    datanew=cat(1,datanew,data(k,:));
                end
            end
        end
    end
end
% plot the data as a coloured density plot
figure(1)
clf
smoothhist2D(datanew,5,[900, 900],0,'image');
set(gca,'YDir','normal');
\end{lstlisting}
\fi
\end{document}